\documentclass[12pt]{article}
\usepackage{epsf,latexsym,amsmath,amssymb,amscd}

\newtheorem{theorem}{Theorem}[section]
\newtheorem{lemma}[theorem]{Lemma}

\newcommand{\myfield}{{\mathbb Q}} 

\newcommand{\objects}[1]{{{\rm Ob}\left( {#1} \right)}} 

\newcommand{\isom}{\simeq} 

\newcommand{\from}{\colon}
\newcommand{\ignore}[1]{}

\newcommand{\Hom}{{\rm Hom}}

\newcommand{\cat}[1]{{\Delta_{#1}}}

\newcommand{\cc}{{\cal C}}

\newcommand{\cs}{{\cal S}}

\newcommand{\bool}{{\mathbb B}}

\newcommand{\proof}{{\par\noindent {\bf Proof}\space\space}}
\newcommand{\proofbox}{\begin{flushright}$\Box$\end{flushright}}

\title{Cohomology in Grothendieck Topologies and Lower Bounds
in Boolean Complexity II: A Simple Example}

\author{Joel Friedman\thanks{
        Departments of Computer Science,
        University of British Columbia, Vancouver, BC\ \ V6T 1Z4, CANADA,
	and
        Departments of Mathematics,
        University of British Columbia, Vancouver, BC\ \ V6T 1Z2, CANADA.
        {\tt jf@cs.ubc.ca}, {\tt http://www.math.ubc.ca/\~{}jf}.
        Research supported in part by an NSERC grant.}}

\begin{document}           
\maketitle                 
\begin{abstract}
In a previous paper we have suggested a number of ideas to attack
circuit size complexity with cohomology.  As a simple example,
we take circuits that can only compute the AND of two inputs,
which essentially reduces to SET COVER.  We show a very special
case of the cohomological approach (one particular free category,
using injective and superskyscraper sheaves) gives the linear
programming bound coming from the relaxation of the standard integer
programming reformulation of SET COVER.
\end{abstract}
\section{Introduction}

In \cite{friedman_cohomology} we introduced several techniques that may
prove useful in using cohomology (on Grothendieck topologies) for obtaining
lower bounds on circuit complexity.  In this paper we simplify this problem
to complexity involving only conjunctions of Boolean functions.
We then show that a simple example of a
Grothendieck topology, sheaves, and open sets
lead to the ``linear programming'' bound.
Furthermore we improve a bound in \cite{friedman_cohomology}.

Let $S$ be a set, and denote by $\bool^S=\{0,1\}^S$ the collection
of functions from $S$ to $\bool=\{0,1\}$ (viewing $1$ as TRUE and
$0$ as FALSE).  
By a {\em formal AND measure} we mean a function, $h$,
from $\bool^S$ to the non-negative reals such that
for all $f,g\in\bool^S$ we have
$$
h(f\wedge g)\le h(f)+h(g)
$$
(compare the notion of a {\em formal complexity measure}, e.g., 
see \cite{wegener}).
Given a subset $\cs_1=\{f_1,\ldots,f_r\}$, let ${\rm size}(f)$ be the
minimum number of elements of $\cs_1$ whose conjunction is $f$; we
define ${\rm size}(f)$ to be infinite if $f$ cannot be expressed as
such a conjuction.  
Equivalently ${\rm size}(f)$ is the size of the smallest formula
computing $f$ via conjunctions of elements of $\cs_1$.
By induction on ``size'' we see that
for any formal AND measure, $h$, we have
$$
{\rm size}(f)\ge h(f)/M,\qquad \mbox{where $M=\max_i h(f_i)$.}
$$
If $h$ also satisfies
$$
h(f)=h(\neg f),
$$
then similarly we have an $h(f)/M$ lower bound on the size of a formula
as before, but where the operations are either conjunctions or the
negation of a conjuction;  then $\log_2(h(f)/M)$ would bound the
formula (or circuit) depth.

Given $S$ and $f_1,\ldots,f_r$, determining ${\rm size}(f)$ is 
NP-complete.  It can be approximated (to within $O(\log r)$) by a
linear program.  We shall show that the ``virtual zero extensions''
described in \cite{friedman_cohomology}, just in the special case
of a free category with sheaves with no non-trivial higher cohomology,
give this linear programming bound (or more precisely its dual).
We finish by recalling the notion of virtual zero extensions.

We sketch the ideas, referring to \cite{friedman_cohomology} for the details.
Let $\cc$ be a finite category, and endow it with the {\em grossi\`ere}
topology (meaning that a sheaf is the same thing as a presheaf).  
Let $F,G$ sheaves of finite dimensional $\myfield$-vector
spaces on $\cc$, and let $U$ be an open set of $\cc$ (i.e., a sieve),
and $Z$ be a closed set. 
We say that a sheaf, $H$, is a virtual $G_{U,Z}$ if there is an exact
sequence,
$$
0\to G_U\to G_{U\cap Z}\oplus H \to G_Z\to 0,
$$
where the maps $G_U\to G_{U\cap Z}$ and $G_{U\cap Z}\to G_Z$ are
the usual maps (i.e., the identity on the intersection of the supports),
and where $G_A$ denotes $G$ restricted to $A$ and extended by zero outside $A$.
See \cite{friedman_cohomology} for conditions on the existence of $H$;
for a free category (see \cite{friedman_cohomology}) $H$ always exists.
The following theorem is an improvement over the bound in
\cite{friedman_cohomology}.
\begin{theorem}\label{th:improve}
In the situation above, with $H$ a virtual $G_{U,Z}$,
we have
$$
{\rm cc}(F,G_{U\cap \overline{Z}}) \le {\rm cc}(F,G)+{\rm cc}(F,G_U)
+{\rm cc}(F,G_{\overline{Z}}),
$$
where ${\rm cc}(A,B)$ is the sum of the dimensions of ${\rm Ext}^i_{\cc}(A,B)$
over all non-negative $i$, and $\overline{Z}$ is the complement of $Z$.
\end{theorem}
We shall prove this theorem in Section~\ref{se:proof}.
Now assume that we have a map $f\mapsto U_f$ from $\bool^S$ to open sets
of $\cc$ such that $U_{f\wedge g}=U_f\cap U_g$.  Then
$$
h(f)={\rm cc}(F,G_{U_f})
$$
is a formal AND measure, provided that ${\rm cc}(F,G)=0$.

The idea to test the ideas of \cite{friedman_cohomology}
on AND's alone arose in conversations with Les Valiant.  We wish to
thank him, as well as Janos Simon, for discussions.
 
\section{AND complexity}


Given $S$ and $\cs_1\subset\bool^S$,
determining ${\rm size}(f)$ is NP-complete,
as it is almost a reformulation of SET COVER (see \cite{vazirani}, for
example, for SET COVER); 
indeed, to determine how many $\cs_1$ elements we need to
obtain $f$, we may assume $f\le f_i$ for all $i$, and then the question
is how many $f_i^{-1}(0)$ are required to cover $f^{-1}(0)$.
The point of this paper is to show that the dual of the usual
``linear programming'' lower bound on size/depth complexity arises
as a very special (and degenerate) case of the sheaf bound.
Specifically, the size complexity is given by the integer program
\begin{eqnarray*}
\min\;\; \sum_{i\in R}\mu_i,&&\-\mbox{subject to} \\
\sum_{i\in R}\mu_i (1-f_i(s)) \ge 1,&&
\-\forall s\in f^{-1}(0) \\
\mu_i=0,1 &&\- i\in R, \\
\end{eqnarray*}
where $R$ is the set of $i$ with $f\le f_i$.
A lower bound to this program is given by the ``relaxed''
linear program where the $\mu_i$
are non-negative reals.  The gap between the integer and linear program
is known to be as high as $O(\log r)$ in certain cases, and never higher
(see \cite{vazirani}).
Equivalent to the linear program is its dual,
\begin{eqnarray*}
\max\;\; \sum_{s\in f^{-1}(0)}\alpha_s,&&\-\mbox{subject to} \\
\sum_{s\in f^{-1}(0)}\alpha_s (1-f_i(s))\le 1,&&
\-\forall i\in R\\
\alpha_s \ge 0 && \- \forall s\in S\\
\end{eqnarray*}

Say that $s\in f^{-1}(0)$ {\em demands $f_i$} if $f_i(s)=0$ and
$f_j(s)=1$ for $j\ne i$.  If for each $i=1,\ldots,r$ there is an
$s_i$ that demands $f_i$, then we take $\alpha_s$ to be $1$ or $0$
according to whether or not $s$ is one of the $s_i$, and then we
see that the LP-bound is exact, i.e., gives the true size complexity
(namely $r$).

For example, consider the case where
$S=\{0,1\}^n$ and the size one functions are
$\{0,1,x_1,\neg x_1,\ldots,\neg x_n\}$; given $f$, we discard
$0,1$ and all size one functions not $\ge f$; we see that either
$f$ is the conjunction of the functions leftover, and the LP-bound
is exact, or $f$ is of infinite size complexity, and the LP-bound
is also infinite (the primal is infeasible, and for the dual
there is an $s$ with
$f(s)=0$ but $f_i(s)=1$ for all leftover $f_i$; then $\alpha_s$
can be taken arbitrarily large).

\section{Improved Inequality}
\label{se:proof}
In this section we prove Theorem~\ref{th:improve}.

The short exact sequence
$$
0\to G_{U\cap \overline{Z}} \to G_U \xrightarrow{\beta} G_{U\cap Z}\to 0
$$
gives a long exact sequence, yielding
\begin{equation}\label{eq:dim_sum}
\dim({\rm Ext}^i(F,G_{U\cap \overline{Z}})) =
\dim({\rm Coker}\,\beta^{i-1}) + \dim({\rm Ker}\,\beta^i),
\end{equation}
where $\beta^i\from {\rm Ext}^i(F,G_U)\to{\rm Ext}^i(F,G_{U\cap Z})$
are the maps arising from $\beta$.
It suffices to bound the right-hand-side of equation~(\ref{eq:dim_sum}).

The virtual zero extension gives a short sequence
$$
0\to G_U\xrightarrow{\beta\oplus\gamma} G_{U\cap Z}\oplus H\to G_Z\to 0.
$$
Letting $\gamma^i$ be, as before, the map $\gamma$ in
${\rm Ext}^i(F,\;\cdot\;)$ gives
$$
\dim({\rm Ext}^i(F,G_{Z})) =
\dim({\rm Coker}\,(\beta^{i}\oplus\gamma^i)) + \dim({\rm Ker}\,
(\beta^{i+1}\oplus
\gamma^{i+1})).
$$
Since ${\rm Coker}\,\beta^i$ injects into ${\rm Coker}\,(\beta^i\oplus
\gamma^i)$, we have
$$
\sum \dim({\rm Coker}\,\beta^i) \le
\sum \dim( {\rm Coker}\,(\beta^i\oplus
\gamma^i) ) \le {\rm cc}(F,G_Z).
$$
And clearly
$$
\sum \dim({\rm Ker}\,\beta^i) \le \sum \dim( {\rm domain}(\beta^i))
= \sum\dim( {\rm Ext^i(F,G_U)} )={\rm cc}(F,G_U).
$$
Summing in equation~(\ref{eq:dim_sum}) yields
$$
{\rm cc}(F,G_{U\cap\overline{Z}}) \le {\rm cc}(F,G_Z)+{\rm cc}(F,G_U).
$$
We now finish the proof using
$$
{\rm cc}(F,G_Z)\le {\rm cc}(F,G)+
{\rm cc}(F,G_{\overline{Z}})
$$
that follows from the exact sequence
$$
0\to G_{\overline{Z}}\to G\to G_Z\to 0.
$$
\proofbox

\section{A Trivial Bound}





In this section we use the notation of \cite{friedman_cohomology}:
if $\cc$ is a category and $P\in\objects{\cc}$, then $k_P$ is the inclusion
$\cat{0}\to\cc$ where $\cat{0}$ is the category with one object, $0$,
and one morphism, and $k_P(0)=P$; also, if $u\from\cc\to\cc'$ is a functor,
$u^*$ is the pullback and $u_*$ (respectively $u_!$) is its right
(respectively left) adjoint (this notation comes 
\cite{sga4.1}, Expos\'e I, Section 5.1).

\begin{lemma} Let $\cc$ be the free category on the graph, $G$. 
For a closed inclusion $i\from Z\to \cc$, let $\Hom_Z(P,Q)$ be the
set of all paths in $G$ all of whose vertices except the last lie
outside $Z$; in particular, if $P\in Z$, then $\Hom_Z(P,Q)$ is empty
if $Q\ne P$ and consists of a single element (the zero length path
about $P$) if $Q=P$.
For
any $P\in\objects{\cc}$
we have
that 
\begin{equation}\label{eq:ii}
i_*i^*k_{P*}\myfield \isom \bigoplus_{Q\in Z}
(k_{Q*}\myfield)^{\Hom_Z(P,Q)}.
\end{equation}
\end{lemma}
\proof 
Let $F_L,F_R$ denote the sheaves on the left- and right-hand-side
of equation~(\ref{eq:ii}).
First note that $i_*i^*$ is simply restriction to $Z$ followed by
extension by $0$, and $F_R$ clearly vanishes outside $Z$.
So it suffices to give anisomorphism
$F_L(X)\isom F_R(X)$ for each $X\in Z$
that is functorial in $X$.

We have that $(k_{P*}\myfield)(X)$ is $\Hom(P,X)$ copies of $\myfield$,
and $\Hom(P,X)$ is the number of paths in $G$ from $P$ to $X$.
For each path in $G$ from $P$ to $X$, once a vertex in
the path, $Q$, falls in $Z$, all subsequent vertices remain in $Z$.
Hence we have a set theoretic bijection
$$
b_X\from \bigcup_Q \Hom_Z(P,Q)\times\Hom(Q,X)\; \to \Hom(P,X)
$$
for each $X$.  Furthermore this bijection is functorial in $X$, in
that if $\phi\from X_1\to X_2$ is a morphism, then we have that
$\phi b_{X_1} = b_{X_2} \phi$, the second $\phi$ acting on each
$\Hom(Q,X_1)$.
This gives the desired functorial isomorphism $F_L(X)\isom F_R(X)$.
\proofbox

Now let $S$ be a finite set, and let
$M_S$ be the graph whose vertices 
are the functions $S\to \{0,1\}$ with one or zero edges from $f$ to $g$
according to whether or not $f\le g$ and $f(s)=g(s)$ for all but exactly
one $s\in S$.
Let $\cc$ be the free graph on $M_S$.  We call $M_S$ the {\em monotone
$S$-cube}, and $\cc$ the {\em $S$-path category}.
For $f\in\objects{\cc}$, let $U_f$ be the smallest open set containing
$f$, i.e., the set of all objects no greater than $f$.
By a {\em subcube} of $\cc$ be mean a collection of objects whose
values at a subset of $S$ are fixed.

Consider the model, $f\mapsto (\cc,F,G_{U_f})$,
where $G=k_{P*}\myfield$
and where $F$ and $P$ are to be specified later.
Since $\cc$ is free, virtual zero extensions always exist.  If
$Z_f$ denotes the complement of $U_f$, we have
$$
{\rm cc}(f)= \sum_{Q\in Z_f} (\dim F(Q))\; |\Hom_{Z_f}(P,Q)|.
$$
Of course, the $\dim F(Q)$ can  be arbitrary non-negative integers by
taking $F$ to be a sum of superskyscraper sheaves, i.e., a sheaf, $F$, where
for each morphism $\phi\in\cc$ we have $F\phi$ is the zero morphism.

We claim this recovers the linear programming bound.  Indeed,
let $P=0$, and let $F$ vanish outside of the points $\delta_s$ where
$s\in S$ and $\delta_s$ is the Dirac delta function at $s$.
Then ${\rm cc}(F,G)=0$, since $F(0)=0$, and
$$
{\rm cc}(g)=\sum_{s} A_s (1-g(s)),
$$
where $A_s$ is the dimension of $F(\delta_s)$.  So we get
$$
{\rm size}(f) \ge \sum_s A_s/M,
$$
for any $M$ with
$$
\sum_s A_s (1-f_i(s)) \le M;
$$
this is just the linear programming bound (restricted to $\alpha_s=A_s/M$
rational, $\alpha_s$ as in Section~2).
Furthermore, it is not hard to see that varying $G$ injective
and $F$ arbitrary and taking
linear combinations we cannot get anything better than the linear
programming bound.  Indeed, since $f=0$, we have ${\rm cc}(f)=0$ unless
$P=0$; so any $P\ne 0$ terms can be discarded in an optimal bound.
Similarly, if $P=Q$, then ${\rm cc}(F,G)$ is already $1$ and ${\rm cc}(f)$ is
no greater, so such terms can be discarded (since the bound ${\rm size}(f)\ge
1$ can always be achieved by the linear program).
Finally, if $P=0$ and $Q\ne 0$ and $Q\ne\delta_S$, then
$$
{\rm cc}(f)=\sum_{(A,B)\in E_S} N_{P,Q}(A,B){\rm cc}_{A,B}(f),
$$
where $E_S$ are the edges of $G_S$,
$N_{P,Q}(A,B)$ is the number of paths from $P$ to $Q$ in $G_S$ that pass
through the edge $(A,B)$, and ${\rm cc}_{A,B}$ denotes the cohomological
complexity when $P=A$ and $F$ is zero outside $B$ and $\myfield$ on $B$.
In the above displayed sum we may discard the $A$ with $A\ne 0$, as mentioned
before, and we are left with a cohomological complexity as before.







The bound we get on ${\rm size}(f)$ can, of course,
be derived without cohomology.  Indeed, consider any formal function
\begin{equation}\label{eq:formal}
h(f)= \sum_{\phi\in\Hom_{Z_f}(P,\;\cdot\;)} A(\phi),
\end{equation}
where $A$ is a any non-negative function.  Then $h$ is a formal AND measure,
since $\Hom_{Z_{f\wedge g}}\subset\Hom_{Z_f}\cup\Hom_{Z_g}$.

We pause for a mild generalization of this notion.
By a {\em conjuctively closed family} we mean an $C\subset \bool^S$
such that $f,g\in C$ implies $f\wedge g\in C$; by a {\em conjunctively
closed complement} we mean the complement in $\bool^S$ of a conjunctively
closed family, or equivalently a $B\subset\bool^S$ such that
$f\wedge g\in B$ implies at least one of $f,g$ lies in $B$.
For such a $B$, we have $\chi_B$, the characteristic function of $B$
is a formal AND measure,
and therefore so is any
non-negative linear combination of such characteristic functions;
this is an essential generalization of the $h$ in
equation~(\ref{eq:formal}).

Let us mention that taking
$G=(k_{P*}\myfield)_U$ for an open set, $U$, yields such a
bound; indeed, it is not
hard to see that we get ${\rm cc}(f)$
as in equation~(\ref{eq:formal}) with $A(\phi)=\beta(t\phi)$ with
$$
\beta(Q)=\dim F(Q) + \dim V(Q) - 2\; {\rm rank}(M),
$$
where
$$
V(Q) = \bigoplus_{\chi\in\Hom_{\overline{U}}(Q,\;\cdot\;)} F(t\chi),
$$
and $M\from V(Q)\to F(Q)$ is the map whose $\chi$ component is
$F(\chi)$; clearly $\beta$ (and therefore $A$) is non-negative.


\end{document}